\documentclass[prl,twocolumn,amsmath,amssymb,superscriptaddress]{revtex4}
\usepackage{graphicx}
\usepackage{dcolumn}
\usepackage{bm}
\begin{document}

\title{Strong Enhancement of Rashba spin-orbit coupling with increasing
anisotropy in the Fock-Darwin states of a quantum dot}
\author{Siranush Avetisyan}
\affiliation{Department of Physics and Astronomy,
University of Manitoba, Winnipeg, Canada R3T 2N2}
\author{Pekka Pietil\"ainen}
\affiliation{Department of Physics/Theoretical Physics,
University of Oulu, Oulu FIN-90014, Finland}
\author{Tapash Chakraborty$^\ddag$}
\affiliation{Department of Physics and Astronomy,
University of Manitoba, Winnipeg, Canada R3T 2N2}

\begin{abstract}
We have investigated the electronic properties of elliptical quantum
dots in a perpendicular external magnetic field, and in the presence
of the Rashba spin-orbit interaction. Our work indicates that the
Fock-Darwin spectra display strong signature of Rashba spin-orbit coupling 
even for a low magnetic field, as the anisotropy of the quantum dot is 
increased. An explanation of this pronounced effect with respect to the 
anisotropy is presented. The strong spin-orbit coupling effect manifests 
itself prominently in the corresponding dipole-allowed optical transitions, 
and hence is susceptible to direct experimental observation.
\end{abstract}

\maketitle

In recent years our interest in understanding the unique effects of
the spin-orbit interaction (SOI) in semiconductor nanostructures 
\cite{halperin_SO} has peaked, largely due to the prospect of the 
possible realization of coherent spin manipulation in spintronic 
devices \cite{spintro}, where the SOI is destined to play a crucial 
role \cite{halperin}. As the SOI couples the orbital motion of the 
charge carriers with their spin state, an all-electrical control of 
spin states in nanoscale semiconductor devices could thus be a reality. 
In this context the Rashba SOI \cite{rashba} has received particular 
attention, largely because in a two-dimensional electron gas the strength 
of the Rashba SOI has already been shown to be tuned by the application 
of an electric field \cite{nitta}. While the earlier studies were 
primarily in a two-dimensional electron gas, the attention has now been 
focused on the role of SOI in a single InAS quantum dot \cite{soi_dot}. 
The quantum dot (QD) \cite{qdbook}, a system of few electrons confined 
in the nanometer region has the main advantage that the shape and size 
of the confinement can be externally controlled, which provides an 
unique opportunity to study the atomic-like properties of these systems 
\cite{qdbook,heitmann}. SO coupling in quantum dots generates anisotropic
spin splitting \cite{haug} which provides important information about
the SO coupling strength.

Extensive theoretical studies of the influence of Rashba SOI in 
circularly symmetric parabolic confinement have already been 
reported earlier \cite{rashba_tc}, where the SO coupling was found 
to manifest itself mainly in multiple level crossings and level 
repulsions. They were attributed to an interplay between the Zeeman 
and the SOI present in the system Hamiltonian. Those effects, in 
particular, the level repulsions were however weak and as a result, 
would require extraordinary efforts to detect the strength of SO
coupling \cite{tunneling} in those systems. Here we show that, by 
introducing anisotropy in the QD, i.e., by breaking the circular 
symmetry of the dot, we can generate a major enhancement of the Rashba 
SO coupling effects in a quantum dot. As shown below, this can be 
observed directly in the Fock-Darwin states of a QD, and therefore 
should be experimentally observable \cite{qdbook,heitmann}. We show 
below that the Rashba SO coupling effects are manifestly strong in an 
elliptical QD \cite{madhav}, which should provide a direct route to 
unambigiously determine (and control) the SO coupling strength. 
It has been proposed recently that the anisotropy of a quantum dot 
can also be tuned by an in-plane magnetic field \cite{nowak}.

The Fock-Darwin energy levels in elliptical QDs subjected to a 
magnetic field was first reported almost two decades ago \cite{madhav}, 
where it was found that the major effect of anisotropy was to lift 
the degeneracies of the single-particle spectrum \cite{elliptic_expt}. 
The starting point of our present study is the stationary Hamiltonian
\begin{eqnarray*}
{\cal H}^{}_{\rm S} &=& \frac1{2m^*}\left({\bf p}-\frac ec{\bf A}^{}_{\rm S}
\right)^2 + V^{}_{\rm conf}(x,y) + {\cal H}^{}_{\rm SO} + {\cal H}^{}_z \\
&=& {\cal H}^{}_0 + {\cal H}^{}_{\rm SO} + {\cal H}^{}_z
\end{eqnarray*}
where the confinement potential is chosen to be of the form
$$V^{}_{\rm conf}=\tfrac12m^*\left(\omega_x^2x^2+\omega_y^2y^2\right),$$
${\cal H}^{}_{\rm SO}=\frac{\alpha}\hbar\left[\mbox{\boldmath${\sigma}$}
\times\left({\bf p}-\frac ec{\bf A}^{}_{\rm S}\right)\right]^{}_z$ is the
Rashba SOI, and ${\cal H}^{}_z$ is the Zeeman contribution. Here $m^*$ 
is the effective mass of the electron, \mbox{\boldmath${\sigma}$} are 
the Pauli matrices, and we choose the symmetric gauge vector potential 
${\bf A}^{}_{\rm S}=\frac12\left(-y,x,0\right)$. As in Ref.~\cite{madhav}, 
we introduce the rotated coordinates and momenta
\begin{eqnarray*}
x &=& q^{}_1\cos\chi - \chi^{}_2p^{}_2\sin\chi, \\
y &=& q^{}_2\cos\chi - \chi^{}_2p^{}_1\sin\chi, \\
p^{}_x &=& p^{}_1\cos\chi + \chi^{}_1q^{}_2\sin\chi, \\
p^{}_y &=& p^{}_2\cos\chi + \chi^{}_1q^{}_1\sin\chi,
\end{eqnarray*}
where
\begin{eqnarray*}
\chi^{}_1 &=&
-\left[\tfrac12\left(\Omega_1^2+\Omega_2^2\right)\right]^{\frac12},\quad
\chi^{}_2 = \chi_1^{-1},\\
\tan2\chi &=& m^*\omega^{}_c\left[2\left(\Omega_1^2+\Omega_2^2\right)
\right]^{\frac12}/\left(\Omega_1^2-\Omega_2^2\right), \\
\Omega_{1,2}^2 &=& {m^*}^2\left(\omega_{x,y}^2+\tfrac14\omega_c^2\right),
\quad \omega^{}_c=eB/m^*c.
\end{eqnarray*}
In terms of the rotated operators introduced above, the Hamiltonian 
${\cal H}^{}_0$ is diagonal
\cite{madhav}
\begin{equation*}
{\cal H}^{}_0=\frac1{2m^*}\sum_{\nu=1,2}\left[\beta^2_\nu p_\nu^2 + 
\gamma_\nu^2q_\nu^2\right],
\end{equation*}
where
\begin{eqnarray*}
\beta_1^2 &=& \frac{\Omega_1^2+3\Omega_2^2+\Omega_3^2}{2\left(\Omega_1^2
+\Omega_2^2 \right)}, \quad \gamma_1^2=\tfrac14\left(3\Omega_1^2+
\Omega_2^2+\Omega_3^2\right),\\
\beta_2^2 &=& \frac{3\Omega_1^2+\Omega_2^2-\Omega_3^2}{2\left(\Omega_1^2
+ \Omega_2^2\right)}, \quad \gamma_2^2=\tfrac14\left(\Omega_1^2+3
\Omega_2^2-\Omega_3^2\right),\\
\Omega_3^2 &=& \left[\left(\Omega_1^2-\Omega_2^2\right)^2+2{m^*}^2
\omega_c^2\left(\Omega_1^2+\Omega_2^2\right)\right]^{\frac12}.
\end{eqnarray*}
Since the operator ${\cal H}^{}_0$ is obviously equivalent to the
Hamiltonian of two independent harmonic oscillators, the states of the
electron can be described by the state vectors $|n^{}_1,n^{}_2;s^{}_z\rangle$.
Here the oscillator quantum numbers $n^{}_i=0,1,2,\ldots$ correspond to the 
orbital motion and $s^{}_z=\pm\frac12$ to the spin orientation of the electron.

\begin{center}
\begin{figure}
\includegraphics[width=8cm]{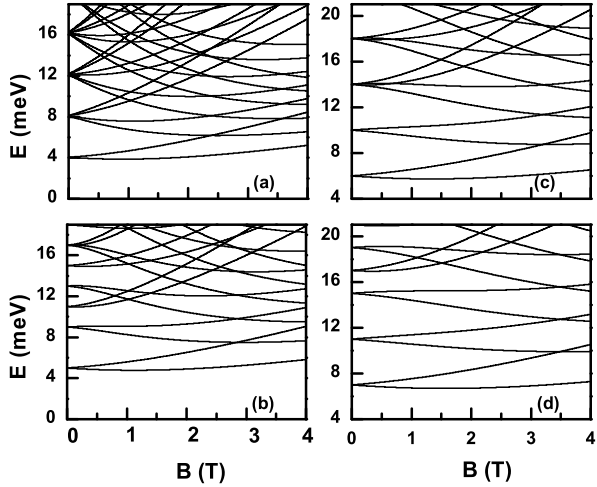}
\caption{\label{fig:alpha=0}  Magnetic field dependence of the low-lying
Fock-Darwin energy levels of an elliptical dot without the Rashba SO 
interaction $(\alpha=0)$. The results are for (a) $\omega^{}_x=4$ meV 
and $\omega^{}_y=4.1$ meV, (b) $\omega^{}_x=4$ meV and $\omega^{}_y=6$ meV, 
(c) $\omega^{}_x=4$ meV and $\omega^{}_y=8$ meV, and (d) $\omega^{}_x=4$ 
meV and $\omega^{}_y=10$ meV.
}
\end{figure}
\end{center}

The Rashba Hamiltonian, in terms of the rotated operators is now written as,
\begin{eqnarray*}
\frac{\hbar}\alpha {\cal H}^{}_{\rm SO}= &&\sigma^{}_x\left(\sin\chi \chi^{}_1
-\cos\chi \omega^{}_0\right)q^{}_1 \\
-&&\sigma^{}_y\left(\sin\chi \chi^{}_1 + \cos\chi \omega^{}_0\right)q^{}_2 \\
-&&\sigma^{}_y\left(\cos\chi - \sin\chi\omega^{}_0\chi^{}_2\right)p^{}_1\\
+&&\sigma^{}_x\left(\cos\chi + \sin\chi\omega^{}_0\chi^{}_2\right)p^{}_2,
\end{eqnarray*}
where $\omega^{}_0=eB/2c$. The effect of the SO coupling is readily handled
by resorting to the standard ladder operator formalism of harmonic oscillators
and by diagonalizing ${\cal H}^{}_\mathrm{SO}$ in the complete basis formed by the
vectors $|n^{}_1,n^{}_2;s^{}_z\rangle$.

The Fock-Darwin states in the absence of the Rashba SOI $(\alpha=0)$ are
shown in Fig.~1, for $\omega^{}_x=4$ meV and $\omega^{}_y=4.1, 6, 8, 10$ meV 
in (a)-(d) respectively. We have considered the parameters of an InAs QD
\cite{rashba_tc} throughout, because in such a narrow-gap semiconductor 
system, the dominant source of the SO interaction is the structural inversion 
asymmetry \cite{zawad}, which leads to the Rashba SO interaction. As expected,
breaking of circular symmetry in the dot results in lifting of degeneracies 
at $B=0$, that is otherwise present in a circular dot 
\cite{madhav,elliptic_expt}. In Fig.~1 (a), the QD is very close to being 
circularly symmetric, and as a consequence, the splittings of the zero-field 
levels are vanishingly small. As the anisotropy of the QD is increased 
[(b) -- (d)], splitting of the levels becomes more appreciable. 

As the SO term is linear in the position and momentum operators it is also 
linear in the raising and lowering ladder operators. It is also off-diagonal 
in the quantum number $s^{}_z$. As a consequence, the SOI can mix only states 
which differ in the spin orientation, and differ by 1 either in the quntum 
number $n^{}_1$ or in $n^{}_2$ but not in both. In the case of rotationally 
symmetric confinements these selection rules translate to the conservation 
of the total angular momentum $j=m+s^{}_z$ in the planar motion of the 
electron.

\begin{center}
\begin{figure}
\includegraphics[width=8cm]{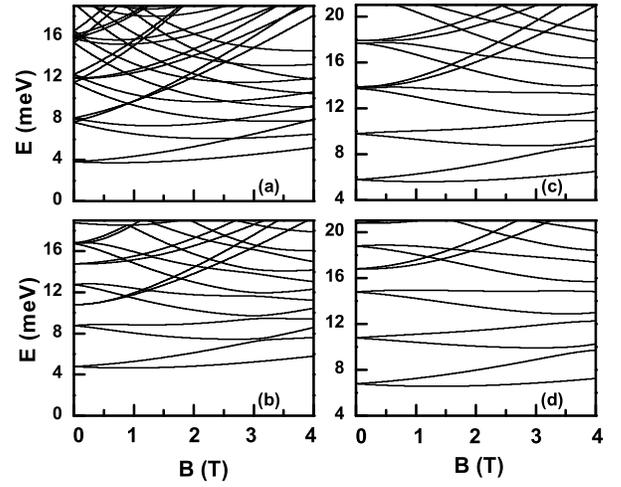}
\caption{\label{fig:alpha=20} Same as in Fig.~\ref{fig:alpha=0}, but for
$\alpha=20$.}
\end{figure}
\end{center}

At the field $B=0$ the ground states $|0,0;\pm\frac12\rangle$ are two-fold
degenerate. Due to the selection rules, this degeneracy cannot be lifted
either by the eccentricity of the dot or by the Rashba coupling. Many of the
excited states, such as $|n^{}_1,n^{}_2;\pm\frac12\rangle$ retain their
degeneracy no matter how strong the SO coupling is or how eccentric the
dot is, as we can see in the Figs.~1-3. At the same time, many other
degeneracies are removed by squeezing or streching the dot. At non-zero
magnetic fields some of the crossings of the energy spectra are turned
to anti-crossings by the Rashba term in the Hamiltonian. For example, 
the second and third excited states in Fig.~2 (a) -- Fig.~2 (d) are 
composed mainly of the states $|0,0;\frac12\rangle$ and $|1,0;-\frac12\rangle$
which are mixed by the ${\cal H}^{}_\mathrm{SO}$ around $B=3\,\mathrm T$ 
causing a level repulsion. We can also see that the squeezing of the dot 
enhances the SO coupling. This can be thought of as a consequence of pushing 
some states out of the way, just as in our example of the state $|1,1;\frac12
\rangle$. SOI mixes it with the state $|1,0;-\frac12\rangle$ causing the
latter state to shift downward in energy thereby reducing the anti-crossing
gap. Squeezing the dot, however moves the state energetically farther away
from $|1,0;-\frac12\rangle$ and so weakens this gap reduction effect.
It is abundantly clear from the features revealed in the energy 
spectra that for a combination of strong anisotropy of the dot and higher 
values of the SO coupling strength, large anti-crossing gaps would appear 
even for relatively low magnetic fields.

The effects of anisotropy and spin-orbit interaction on the energy spectra 
above are also reflected in the optical absorption spectra. Let us turn our 
attention on the absorption spectra for transitions from the ground state 
to the excited states. For that purpose we subject the dot to
the radiation field
$$\mathbf A^{}_\mathrm R=A^{}_0\hat{\bm\epsilon}\left(
\mathrm e^{\mathrm i(\omega/c)´\hat{\mathbf n}\cdot\mathbf r
-\mathrm i\omega t}
+\mathrm e^{-\mathrm i(\omega/c)´\hat{\mathbf n}\cdot\mathbf r
+\mathrm i\omega t}\right),$$
where $\hat{\bm\epsilon}$, $\omega$ and $\hat{\mathbf n}$
are the polarization, frequency and the direction of propagation
of the incident light, respectively. We let the radiation enter the dot along
the direction perpendicular to the motion of the electron, that is parallel
to the $z$-axis. Due to the transversality condition the polarization vector
will then lie in the $xy$-plane.

\begin{center}
\begin{figure}
\includegraphics[width=8cm]{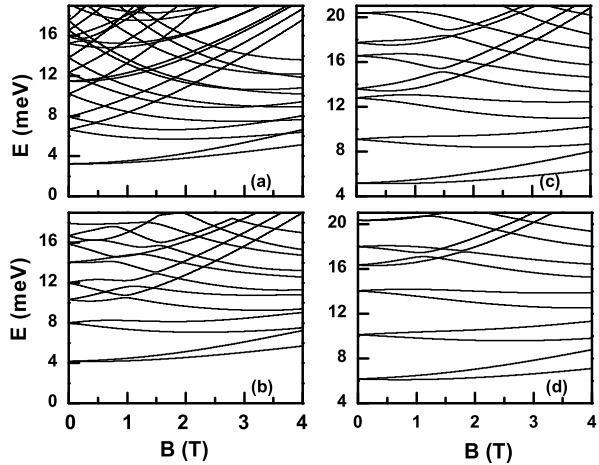}
\caption{\label{fig:alpha=40} Same as in Fig.~\ref{fig:alpha=0}, but for
$\alpha=40$.}
\end{figure}
\end{center}

As usual, we shall make two approximations. First we assume the intensity
of the field be so weak that only the terms linear in $\mathbf A^{}_\mathrm R$
has to be taken into account. Then the effect of the radiative magnetic
field on the spin can be neglected as well. So we can simply replace in the 
stationary Hamiltonian ${\cal H}^{}_\mathrm S$ the vector potential 
$\mathbf A^{}_\mathrm S$ with the field $\mathbf A=\mathbf A^{}_\mathrm S 
+ \mathbf A^{}_\mathrm R$. Discarding terms higher than linear order in 
$\mathbf A^{}_\mathrm R$ leads to the total Hamiltonian
$${\cal H}={\cal H}^{}_\mathrm S+{\cal H}^{}_\mathrm R,$$
where the radiative part ${\cal H}^{}_\mathrm R$ is given by
$${\cal H}^{}_\mathrm R =-\frac e{m^{}_ec}\,\mathbf A^{}_\mathrm R\cdot
\left(\mathbf p-\frac ec\mathbf A^{}_\mathrm S\right)-\frac{\alpha e}{\hbar c}
\, \left[\bm\sigma\times\mathbf A^{}_\mathrm R\right]^{}_z.$$
The radiative Hamiltonian, even in the presence of the Rashba SO coupling can 
be expressed in the well-known form
$${\cal H}^{}_\mathrm R=\mathrm i\frac e{c\hbar}\mathbf A^{}_\mathrm R\cdot
\left[\mathbf x,{\cal H}^{}_\mathrm S\right], $$
$\mathbf x$ being the position operator in the $xy$-plane.

\begin{center}
\begin{figure}
\includegraphics[width=7cm]{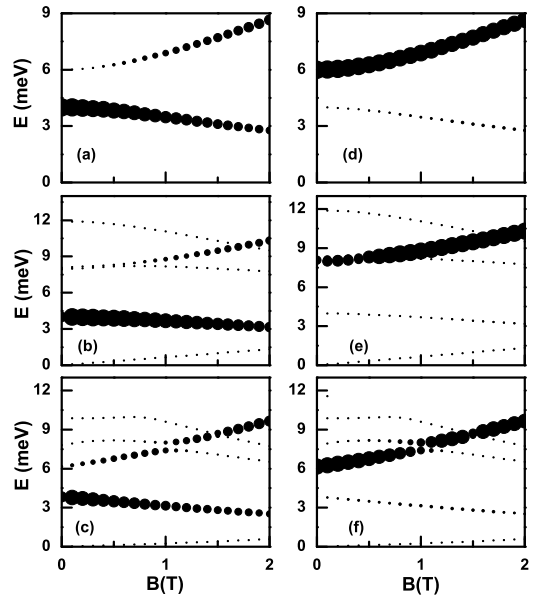}
\caption{\label{fig:trans1} Optical absorption (dipole allowed) specta 
of elliptical QDs for various choice of parameters: (a) i$\alpha=0$, 
$\omega^{}_x=4$ meV, $\omega^{}_y=6$, (b) $\alpha=20$, $\omega^{}_x=4$ meV, 
$\omega^{}_y=8$ meV, and (c) $\alpha=40$, $\omega^{}_x=4$, $\omega^{}_y=6$. 
The polarization of the incident radiation is along the $x$-axis. The 
parameters for (d)-(f) are the same, except that the incident radiation 
is polarized along the $y$-axis. The areas of the filled circles are 
proportional to the calculated absorption cross-section.
}
\end{figure}
\end{center}

Our second approximation is the familiar dipole approximation. We assume 
that the amplitude of radiation can be taken as constant within the quantum 
dot, so that we are allowed to write the field as
$$\mathbf A^{}_\mathrm R\approx A^{}_0\hat{\bm\epsilon}\left(
\mathrm e^{-\mathrm i\omega t}
+\mathrm e^{\mathrm i\omega t} \right). $$
Since the transition energies expressed in terms of radiation frequences
are of the order of THz, the corresponding wavelengths are much larger
than the typical size of a dot, thus justifying our approximation. Applying 
now the Fermi Golden Rule leads to the dipole approximation form
$$ \sigma^{}_\mathrm{abs}(\omega)=4\pi^2\alpha^{}_\mathrm f\omega^{}_{ni}
\left|\langle n|\hat{\bm\epsilon}\cdot\mathbf x|i\rangle\right|^2
\delta(\omega^{}_{ni}-\omega) $$
of the absorption cross section for transitions from the inital state 
$|i\rangle$ to the final state $|n\rangle$. Here $\alpha^{}_\mathrm f$ 
is the fine structure constant and $\omega^{}_{ni}$ is the frequency 
corresponding to the transition energy $\hbar\omega$.

The familiar dipole selection rules for oscillator states dictate largely 
the features seen in Fig.~4. In the absence of the SOI, these rules 
-- the spin state is preserved and either $n^{}_1$ or $n^{}_2$ is changed 
by unity -- completely determine the allowed two transitions $\left|0,0;-
\frac12\right\rangle \rightarrow \left|1,0;-\frac12\right\rangle$ and
$\left|0,0;-\frac12\right\rangle\rightarrow \left|0,1;-\frac12\right\rangle$. 
In contrast to the case of circular dots the absorption in the elliptical dot 
depends strongly on the polarization. This is explained by noting that the 
oscillator strengths
$$f^{}_{ni}=\frac{2m^*\omega^{}_{ni}}\hbar\left|\left\langle n|
\hat{\bm\epsilon}\cdot\mathbf x\right|i\rangle\right|^2.$$
actually probe the occupations of quantum states related to oscillations 
in the direction of the polarization $\hat{\bm\epsilon}$. In a circular dot 
all oscillation directions are equally probable at all energies implying
that the oscillator strengths are independent of the polarization and depend 
only slightly on the transition energy via $\omega^{}_{ni}$, and the final
state quantum numbers $n^{}_{1,2}$. When the dot is 
squeezed in the $y$-direction, say, the oscillator states related to the 
$y$-motion are pushed up in energy. This means that the polarization being 
along $x$-axis most of the oscillator strength comes from transitions to 
allowed states with lowest energies. Similarly, when the incident radiation 
is polarized along the $y$-axis most of the contribution is due to the 
transitions to the oscillator states pushed up in the energy. In elliptical 
dots the oscillator states are not pure $x$- and $y$-oscillators but their 
superpositions. Therefore in addition to the main absorption lines, other 
allowed final states have also non-vanishing oscillator strength. Furthermore,
as one can see by looking at the phase space rotation formulas the external 
magnetic field tends to rotate directions of the oscillator motion causing 
a shift of the oscillator strength from an allowed transition to another. 
This is exactly what we see in Fig.~4 (a) and Fig.~4 (d).

Even in the presence of the SOI the two allowed final oscillator states 
provide major contributions to the corresponding corrected states. Hence 
we still see two dominant absorption lines. However, now many forbidden 
transitions have become allowed. The lowest absorption line corresponding 
to the transition between Zeeman split states with main components
$\left|0,0;-\frac12\right\rangle$ and $\left|0,0;\frac12\right\rangle$
provides a typical example. The transition involves a spin flip and is 
therefore strongly forbidden without the SOI. Because the SOI mixes the 
state $\left|1,0;\frac12\right\rangle$ into the former one and the 
$\left|0,1;-\frac12\right\rangle$ into the latter one, the transition 
becomes possible. The appearance of other new lines can be explained by 
analogous arguments. There are also additional features involving
discontinuities and anti-crossings in Fig.~4. A comparision with the energy 
spectra indicates that these are the consequences of the anti-crossings 
present in the energy spectra.

It is also readily verified that the oscillator strengths satisfy
the Thomas-Reiche-Kuhn sum rule \cite{TRK}
$$\sum_nf^{}_{ni}=1.$$
In terms of the cross section this translates to the condition
$$\int_{-\infty}^\infty\sigma^{}_\mathrm{abs}(\omega)\,\mathrm d\omega
=\frac{2\pi^2\hbar\alpha^{}_{\mathrm f}}{m^*}.$$
The absorptions visible in Fig.~4 practically saturate the sum rule,
the saturation being, of course complete in the absence of the SOI in
panels (a) and (d). The largest fraction (of the order of 1/10)
of the cross section either falling outside of the displayed energy scale
or having too low intensity to be discernible in our pictures is
found at the strongest Rashba coupling in the panels (c) and (f) for
large magnetic fields, as expected.

The results presented here clearly indicate that, the anisotropy of a
QD alone causes lifting of the degeneracies of the Fock-Darwin levels at
B=0, as reported earlier \cite{madhav}. However, for large SO coupling
strengths $\alpha$, the effects of the Rashba SOI, mainly the level
repulsions at finite magnetic fields, are maginified rather significantly
as one introduces anisotropy in the QD. This is reflected also in the
corresponding dipole-allowed optical transitions where the distinct 
anti-crossing behavior is observed that is a  direct manifestation of 
the anti-crossings in the energy spectra. This prominent effect of the 
Rashba SOI predicted here could be confirmed experimentally in optical
spectroscopy and the Fock-darwin spectra of few-electron QDs 
\cite{heitmann,FD_expt,ILS_expt}. It would also provide a very useful 
step to control the SO coupling in nanostructures, en route
to semiconductor spintronics \cite{spintro}.

The work was supported by the Canada Research Chairs Program of the
Government of Canada.

\end{document}